\documentclass[iop, numberedappendix]{./emulateapj}
  \usepackage{natbib}

\def\phs{ph~cm$^{-2}$~s$^{-1}$}
 
  \shortauthors{Jourdain et al.}
\shorttitle{Cygnus X-1 emission  with \textit{INTEGRAL} SPI}

\begin{document}

\title{The emission of Cygnus X-1: observations with \textit{INTEGRAL} SPI \footnote{
Based on observations with INTEGRAL, an ESA project with instruments and science data centre
funded by ESA member states (especially the PI countries: Denmark, France, Germany, Italy, 
Spain, and Switzerland), Czech Republic and Poland with participation of Russia and USA.} from 20 keV to 2 Mev}

\author{ E. Jourdain, J. P. Roques and J. Malzac}
\affil{ Universit\'e de Toulouse; UPS-OMP; IRAP;  Toulouse, France\\
 CNRS; IRAP; 9 Av. colonel Roche, BP 44346, F-31028 Toulouse cedex 4, France}


\author{\it Received  ; accepted  }


\begin{abstract}
 
We report on  Cyg X-1 observations performed  by the SPI telescope onboard the INTEGRAL 
mission and distributed over  more than 6 years.
We investigate the variability of the intensity and spectral shape
of this peculiar source in the hard X-rays domain, and more particularly up to the 
MeV region. 
We first study the total averaged spectrum which presents 
the best signal to noise ratio (4 Ms of data). Then, we   refine  
our results by building mean spectra by periods and gathering those of similar 
hardness.
  
Several spectral shapes are observed with important changes in the curvature between 
20 and 200 keV, even at the same luminosity level.
 In all cases, the emission decreases sharply above
700 keV, with flux values  above 1 MeV (or upper limits)  well  below the recently 
reported polarised flux (Laurent et al. 2011),
 while compatible with the MeV emission detected some years ago  by 
CGRO/COMPTEL (McConnell et al., 2002).
   
 Finally, we take advantage of the spectroscopic capability of the instrument to seek for
 spectral features in the 500 keV region   with  negative results
 for any  significant  annihilation emission on 2 ks and days timescales, as well as in the total dataset. \\
  
\end{abstract}      

\keywords{radiation mechanisms: general--- Gamma-rays:individual (Cyg X-1) ---  gamma rays: observations 
 }

\maketitle

\section{Introduction}
 Cyg X-1 is an unavoidable target for any high energy instrument. Being one of the most luminous
sources (up to the MeV range), it represents an ideal lab to study the mechanism at work 
in the direct environment of a black hole. There, the accretion flow is thought to
form an optically thick disk and/or an optically thin corona, while the observed radio jets  could originate from the same area (see e.g.the review by Done, Gierli{\'n}ski \& Kubota 2007). The high energy radiation provides insights on the physical processes at work in this region. Thanks to its persistent high flux ($\sim$ 1 Crab) and usually hard spectrum,  Cygnus X-1 is easy to observe and there are numerous measurements of its the spectrum and variability in the X-rays. At higher energies however,  the results are very scarce. Very few instruments were able to explore the properties of the emission above 200 keV.
 
The high energy emission of Cyg X-1 is relatively well known from soft X-rays up to a few hundreds  keV. The shape of the spectrum is variable and can change dramatically on time scales as short as a day. There are however two main, relatively stable, spectral states: the Hard State (HS, corona dominated) and the Soft State  (SS, disk dominated). Their description can be found in various papers (see for example Liang \& Nolan 1984;  Gierlinski et al., 1997, 1999).
So far, the MeV region of the spectrum was best explored by the Compton 
Gamma-ray Observatory (CGRO). McConnell et al (2002) have shown
that, even though the HS hard X-ray emission is dominated by a thermal Comptonization
 component, both  HS and SS spectra present a non thermal powerlaw component
extending above 1 MeV with a slope of 2.6 in the SS and  $>\sim$ 3 in the HS.
Moreover, broad features around 1 MeV have been observed in several occasions in the past  
(see for example, Bassani et al. 1989 and references therein).

Since CGRO observatory however, only the \textit{INTEGRAL} mission contains instrument exploring 
the  same energy domain.  
Recently, Laurent et al. (2011), stacked all the \textit{INTEGRAL/IBIS}  data available for
 Cygnus X-1 and detected  the presence  of a non-thermal power law component between 400 keV 
 and 2 MeV. Interestingly they found that contrary to the thermal Comptonization component 
 present below 400 keV, the non-thermal power-law emission appears to be strongly polarized.
This non-thermal component appears to have a flux that is stronger than that
 measured by CGRO by a factor 5-10 and a much harder spectral slope $\Gamma
  \simeq 1.6$.   
Here, we use another \textit{INTEGRAL} instrument operating in this energy range,
the spectrometer SPI, to investigate
the high energy spectral shape of  Cyg X-1 and test
the presence of non-thermal high energy excess.\\
Our first goal is to take advantage of  the sensitivity achieved with the SPI detector 
and the large amount of data and perform a detailed analysis 
of the energy extent of the hard X-ray emission together with its spectral variability.\\
Moreover, the spectroscopic capability of the Germanium crystals allow us to seek for 
the presence of  spectral features linked to the annihilation  process.
 
 Hereafter, we present briefly the instrument, data set and the method followed for the analysis.
Then, we report on our results and start by examining  the total mean spectrum to
determine the  emission  above a few hundreds of keV, where scarcity of photons
imposes  exposure as long as possible.
In a second step, we analyse the  source behavior during individual revolutions, and   
build several averaged spectra, following some hardness criteria
which can be considered as characteristic of the spectral state of the source.
We conclude with a comparison with previous results. 
 
\section{Instrument, observations and data analysis}
SPI is a spectrometer aboard the  INTEGRAL observatory operating between 20 kev and 8 MeV.
 The description of the instrument
and its performance can be found in Vedrenne et al. (2003) and Roques et al. (2003).
The main features of interest for our study are a large FoV ($30^\circ$) with an 
angular resolution of $2.6^\circ$ (FWHM) based on a coded aperture mask.
The Germanium camera, beyond an excellent 
spectroscopic capability, offers a good sensitivity  
over more than 2 decades in energy with a unique instrument.  It is surrounded by 
a 5-cm thick BGO shield  (ACS, Anti-Coincidence Shield) which measures 
the particle flux. This latter can be used as a good tracers of the background level.
\\
During a 3-day orbit,  the usual dithering strategy (Jensen et al., 2003) consists of a hundred of 30-40 min exposures
(also called scw for 'science window'),
with a given region scanned by  $2^\circ$ steps following pre-determined patterns.
The recommended pattern for SPI observations is a grid of 5X5 around the chosen target.
Unfortunately, in order to content more proposers, except a few exceptions, most of the 
Cyg X-1 data has been obtained through 'amalgamated' observations, ie with the
pattern center somewhere between Cyg X-1 and Cyg A region. As a consequence, 
Cyg X-1 appears only in one side of the FoV, reducing the mean efficient area 
(the source is partially coded), with some interruptions in the
observation sequences, when the source goes out from the field of view.
For our analysis, we have selected in the whole INTEGRAL observation plan, those revolutions 
in which Cyg X-1 
is included in the $\pm 13 ^\circ$ FoV during more than 20 scw (50 ks). 
This gives a total of 42 revolutions. These observations encompass
4 Ms of effective time, from 2003, June  to 2009 December. They were 
grouped  together according to temporal proximity into 13 periods.
Exposures with high background level (entry/exit of radiation belts, solar activity) or large source off-angle
(source/pointing axis angle beyond 13$^\circ$) have been removed from our dataset.
Table 1 gives some details about the observation pattern, beginning, end and useful duration, for each
 of the defined periods.
  
We follow the analysis method described in Jourdain \& Roques (2009), 
based on a sky model fitting, through a $\chi^2$ minimization.
This methods makes use of the Pulse Shape Discrimination (PSD) system, a second electronic chain operating in parallel,
 to eliminate spurious events occurring in the MeV region. In the 650 keV-2.2 MeV energy range, we use only the PSD tagged
events, while below 650 keV, all the single events are analysed. This procedure has
been validated with the Crab Nebula observations.\\
In this work, we consider a common sky model for all the observations,
which consists in 4 sources, namely: Cyg X-1, Cyg X-3, EXO2023+375 and GRS1915+105.
To keep the number of degrees of freedom to a minimum, we consider that the three latter have a constant flux over
the revolution timescale. For Cyg X-1, we consider a variability timescale of 1 scw ($\sim$ 2 ks)
for the lightcurve and hardness ratio studies, while the spectra have been built
assuming a constant source flux during each revolution.\\ 
The background determination relies on empty field observations. Because of detector  failures, 
it is important to use proper empty fields for each configuration. Once the uniformity map
of the detector plan is fixed, we allow its normalisation to vary on a timescale of $\sim$ 10-20 ks.\\
The  incident photon spectra  have been deconvolved and fitted 
through the spectral fitting procedure  from {\sc xspec} 11 (Arnaud 1996).
Due to some uncertainties on the energy response in the lowest channels and threshold effects,
we  exclude the  first channels ($E <$ 23.5 keV) from the fit process.
The spectra presented contain thus 41 (more or less logarithmic) channels between 
23.5 keV and 2.2 MeV. Note that  there are two strong narrow background lines  
(10 to 40 times higher than the continuum emission) at 139 and 198 keV. We use very narrow 
channels around these energies to isolate the line emission and prevent contamination 
of the neighbouring channels. They may appear 
as offset data points with large error bars in the spectra.
\section{Study of the spectral shape}

\subsection{The total mean spectrum }
 
\begin{figure}
\plotone{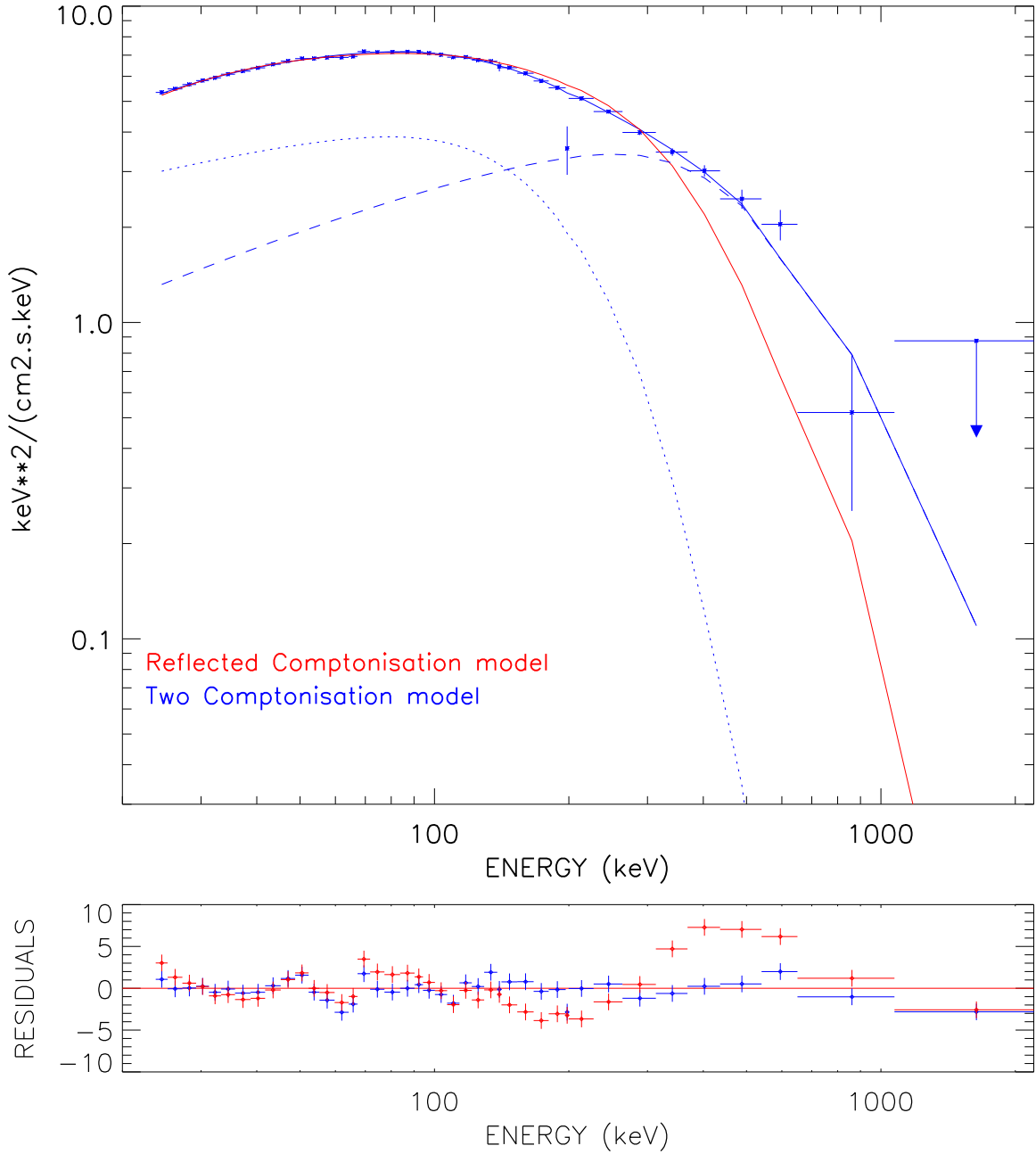} 
\caption{ Cyg X-1 averaged spectra for the whole data set (4 Ms of useful
 duration between 2003 and 2009). Approximations with Comptonization + reflection
 and two Comptonization models are represented by solid lines.
0.5\% systematic errors have been added to the data. Upper limits are at 2$\sigma$ level.
See Table 2 for models parameters.}
\label{fig:tot.comptt1et2}
\end{figure}

The total mean spectrum gathers the whole set of available clean data (4 Ms of observation
distributed over more than 6 years). 
It has been built by extracting one spectrum per revolution then summing them.
 Due to the impressive signal to noise ratio at low energy, we are unable to ensure that 
 the  response matrices are  known with a sufficient precision. We have thus added 0.5\%
 of systematic errors to the data during the fit procedure.\\
 The data have been first fitted  with a simple  analytical cutoff power law and 
a Comptonization + reflection model (reflect*comptt in {\sc xspec}).
Residuals clearly show deviation from these models above 200-300 keV where an excess of emission appears 
above the model prediction. To go further,  we keep  a Comptonization
law (+ its reflected component) in the low energy part and focus on the high energy emission.
Following Laurent et al. (2011), we first try to model this additional component by a single 
power law. The fit converges toward a photon index of $\sim1.8$ (close to the 1.6 $\pm$ 0.2 value
reported by these authors) but the $\chi^2$ value 
remains unacceptable, with a huge contribution of the last channels.
 Indeed, the high energy part doesnot follow such a power-law : instead,
the emission presents a rather sharp decrease after 700 keV and the source is not detected
above 1 MeV.
A better result is obtained when modeling this high energy component by a second Comptonization
region with a temperature  around 100 keV. As several couples (kT, $\tau$) can reproduce the data,
 we fixed $\tau$  to 0.5 and obtained a best fit value of kT=123 keV (see Table 2). 
 Fig. \ref{fig:tot.comptt1et2} presents the observed spectrum with one and two Comptonization models.
Even though the set of parameters proposed in Table 2 is only a possible solution,  
the two Comptonization model provides a  good description of the data and
more specifically  of the curvature observed up to 1 MeV.

\subsection{Flux, hardness  and hardness versus flux evolution}
\begin{figure}
\plotone{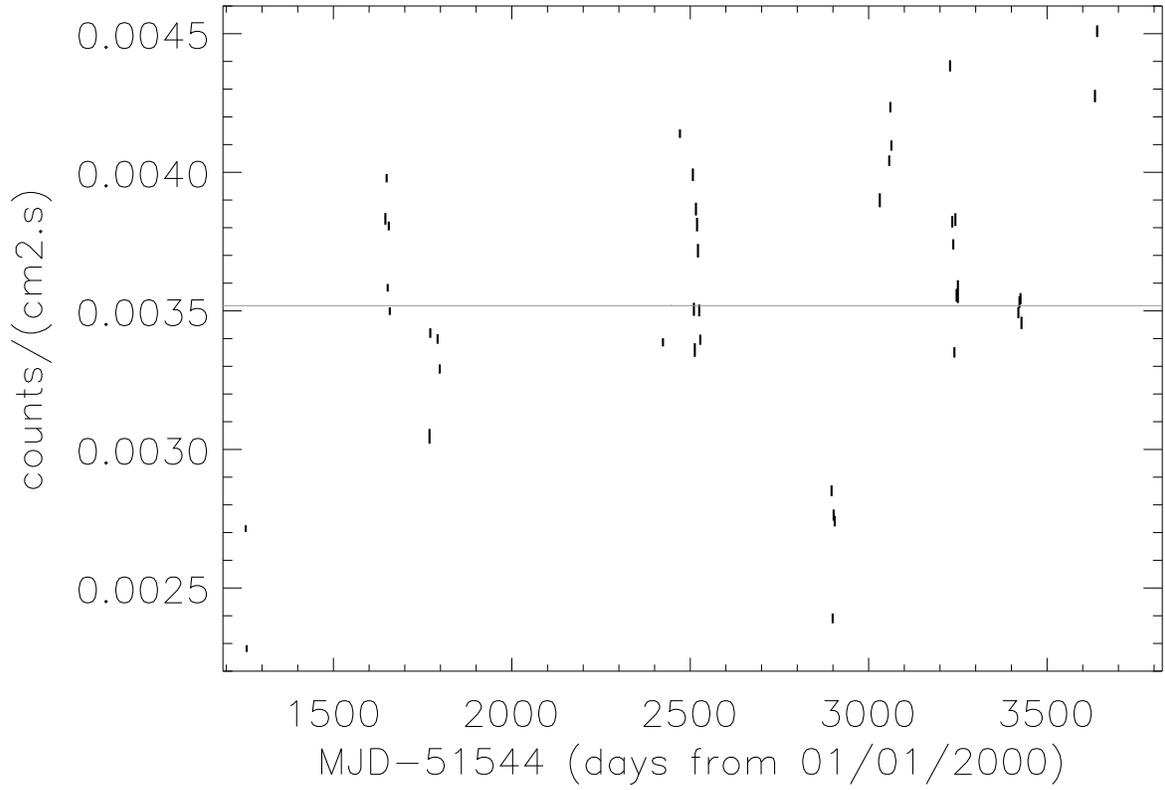}
\caption{Light curves of Cyg X-1 in the 30-70 keV energy range 
 from \textit{INTEGRAL} SPI since
the beginning of the mission. Each point represents a (part of) revolution.
The straight line represents the mean flux.}
\label{fig:clrev}
\end{figure}

\begin{figure}
\plotone{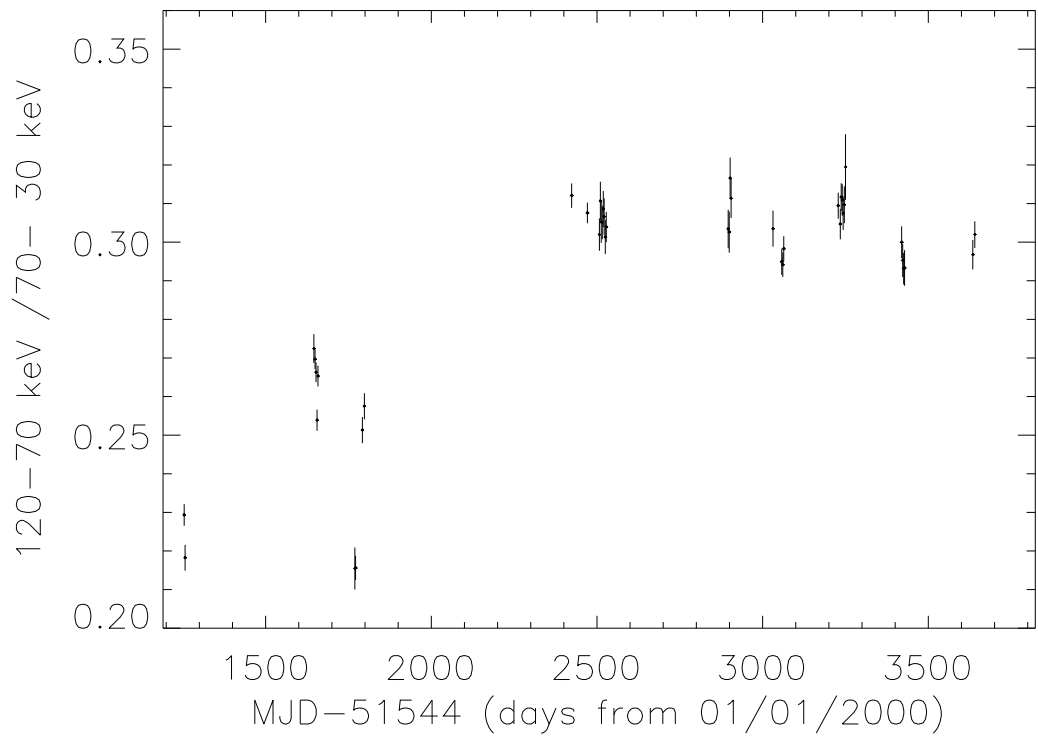}
\caption{Hardness ratio evolution of Cyg X-1. 
Each point represents a (part of) revolution.}
\label{fig:hrrev}
\end{figure}

\begin{figure}
\plotone{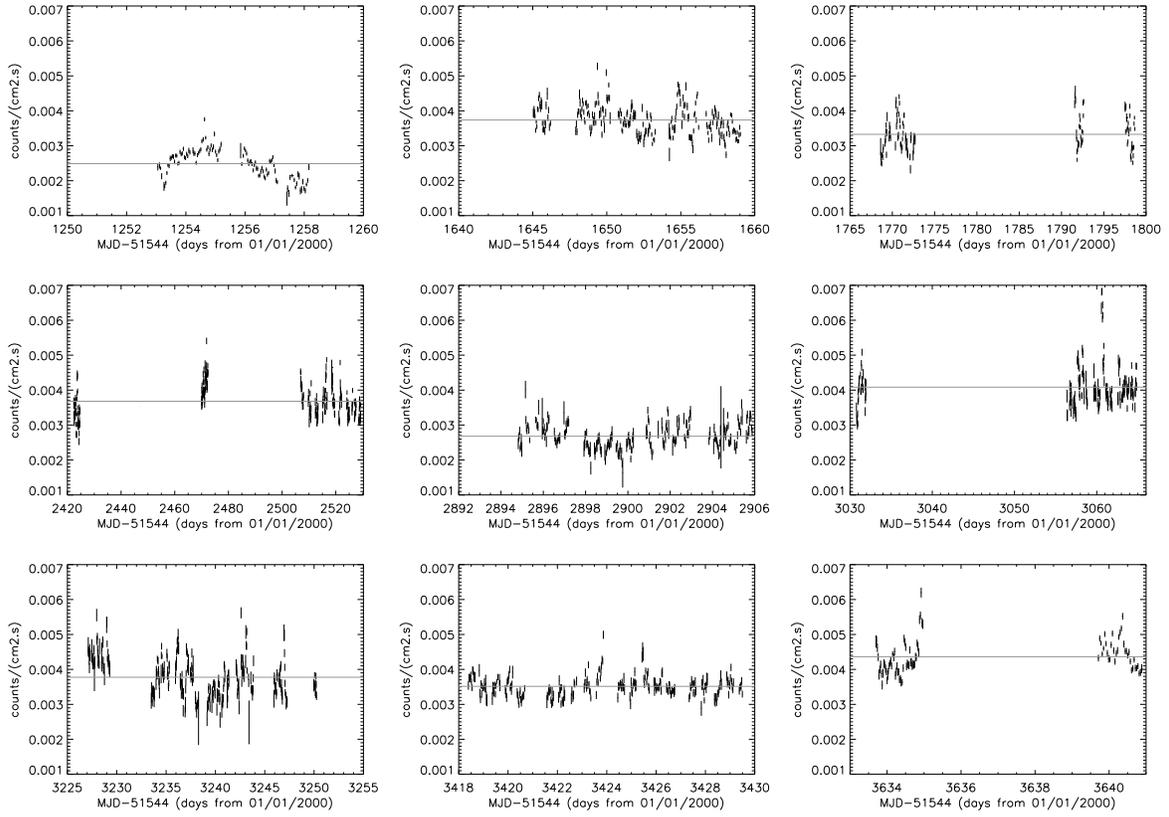}
\caption{The same as Fig. \ref{fig:clrev}, detailed by periods.
 Each point represents a science windows (scw, $\sim$ 2000 s). For each period,
  the straight line represents the mean flux. }
\label{fig:clscw}
\end{figure}

\begin{figure}
\plotone{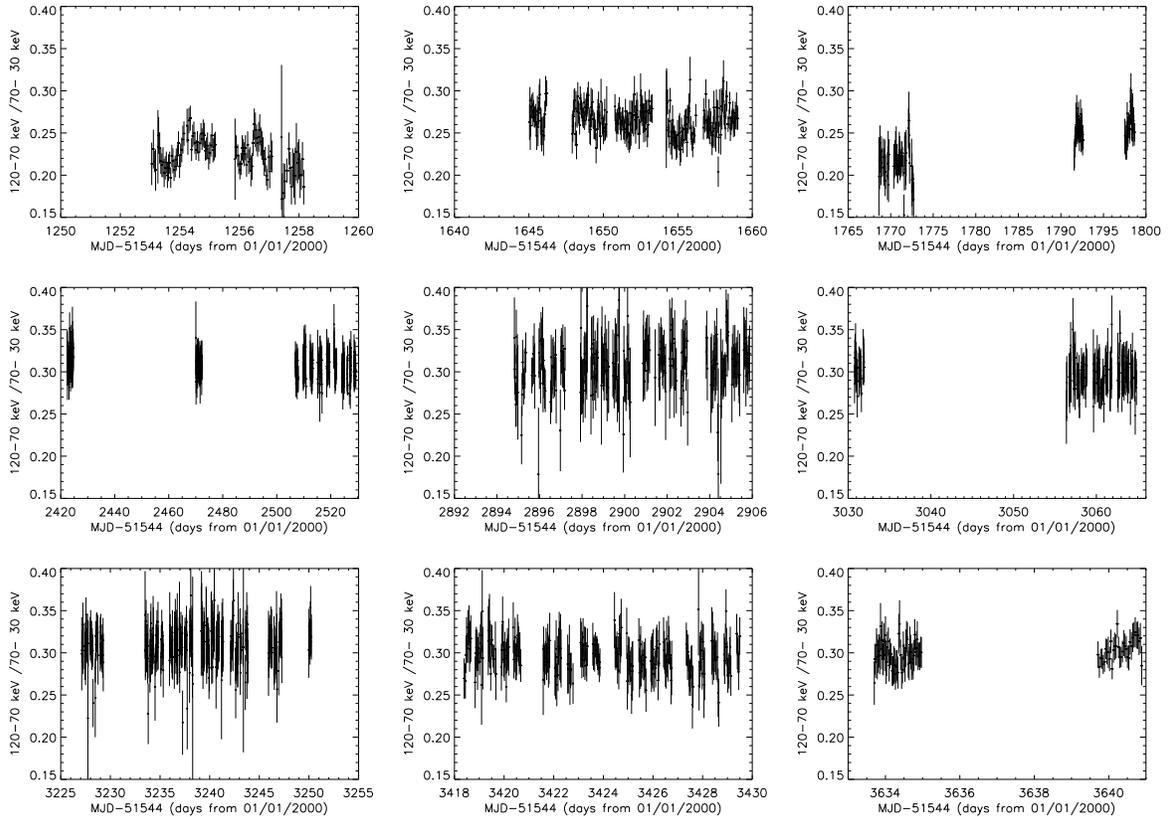}
\caption{The same as Fig. \ref{fig:hrrev}, detailed by periods. 
Each point represents a science windows (scw, $\sim$ 2000 s).}
\label{fig:hrscw}
\end{figure}

The MeV emission may depend on the source spectral state or intensity. To follow potential changes in the source emission, we have 
analysed in details each revolution and used the information to group observations corresponding 
to similar states.\\
Fig. \ref{fig:clrev} displays the temporal evolution of Cyg X-1 in 30-70 keV energy range, 
with the fluxes averaged by revolution (100-200 ks timescale) and
Fig. \ref{fig:hrrev} the evolution of the hardness  
(defined in the 30-120 keV range by $F_{70-120 keV}/F_{30-70 keV}$).
In Fig. \ref{fig:clscw} and Fig. \ref{fig:hrscw}, we present the same flux and hardness  but detailed 
on time scales of a few days  (with a resolution of one scw ie $\sim$ 2000 s).
These graphs illustrate  clearly the variability of the source. 
The hypothesis of constant flux has been tested by $\chi^2$ tests and  is strongly rejected 
for all periods (reduced $\chi^2$ always greater than 5.9 for a number of degrees of 
freedom ranging from 100 to 300). The variability appears to be  chaotic but in
 both cases  over a limited amplitude: the source is always detected and varies by a
  factor not greater than 3.  We can recognize the usual temporal behavior of Cyg X-1
 (see e.g. Ling et al. 1987 and Zdziarski et al. 2002). Note that the first period, which  corresponds the lowest and softest state  in our sample, has been analysed by Malzac et al. (2006). They identified the (rather unusal) source behavior  as a mini or failed transition between soft and hard state, in a so called "intermediate state". However, no robust (that is, lasting more than a few hours) incursion in the soft state can be reported during our observations and we conclude that Cyg X-1 was always in a hard (LH) or intermediate states.

\begin{figure}
\plotone{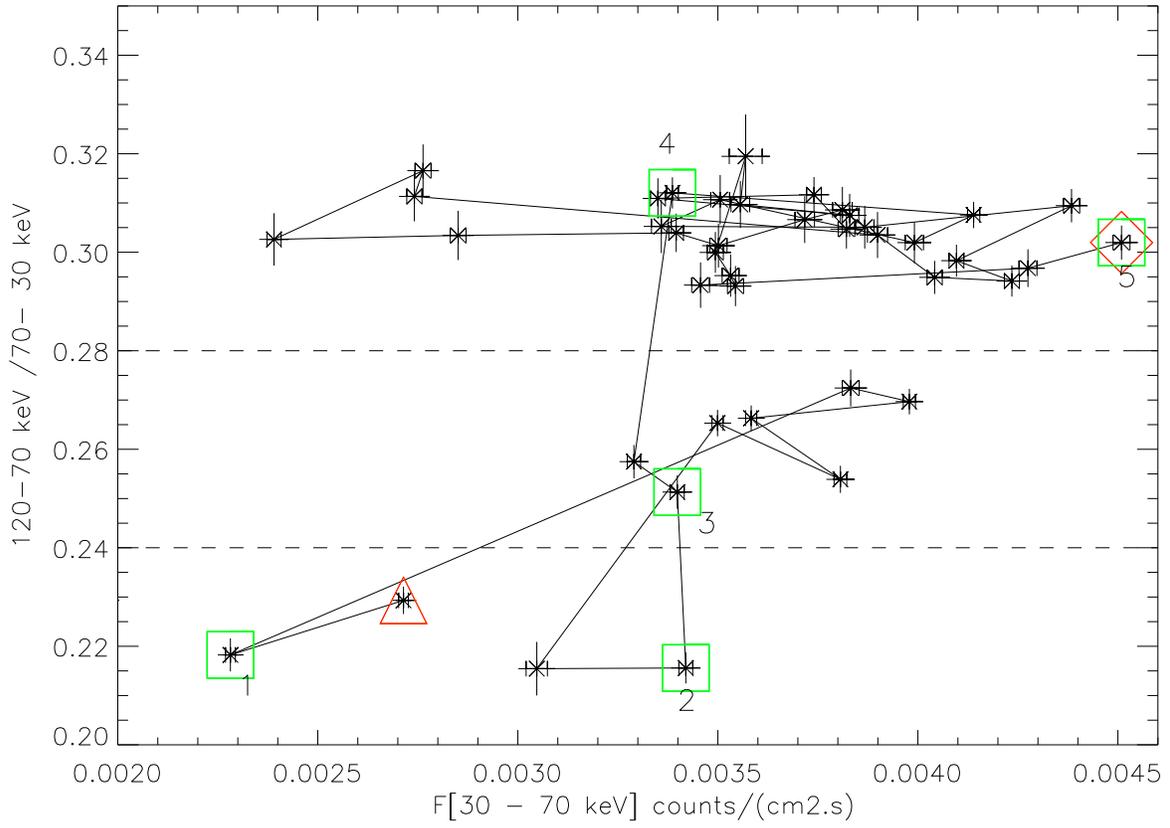}
\caption{Hardness versus flux evolution. Each point represents a (part of) revolution.
The solid straight line joins points in the
chronological order (from the triangle to the diamond symbols). 
The squares with numbers correspond to revolutions used in the next figure.
Dashed lines materialize  the limits used to build different subsets of data (see text
and Figure \ref{fig:3HR}).
}

\label{fig:hrflux}
\end{figure}

 Fig. \ref{fig:hrflux} displays the hardness as a function of the 30-70 keV source flux
 (revolution averaged values). The hardness intensity plane can be divided in 
 three regions corresponding to the main clusters of data points.  Those three
  regions are outlined in Fig.~\ref{fig:hrflux}. The first two regions gather 
   the points with  hardnesses respectively
below 0.24 and  between 0.24 and  0.28, and, incidentally, correspond to the first part
 of the INTEGRAL mission 
(revolutions 79 to 261  ie June 2003 - November 2004).  \\
In the third group, the flux levels span a broader range, extending toward higher values,
but the mean hardnesses never decrease below 0.29. During this period,
which covers more than 3 years (mid 2006 up to end of 2009), the source evolution consists of 
 an intensity variation  within a factor
 of $\sim$ 2  without any visible change in the emission hardness (or spectral shape), 
 so without notable change in the underlying processes.  This  behavior has already been 
 observed in this source and is well known in transient sources.
 
\begin{figure}
\plotone{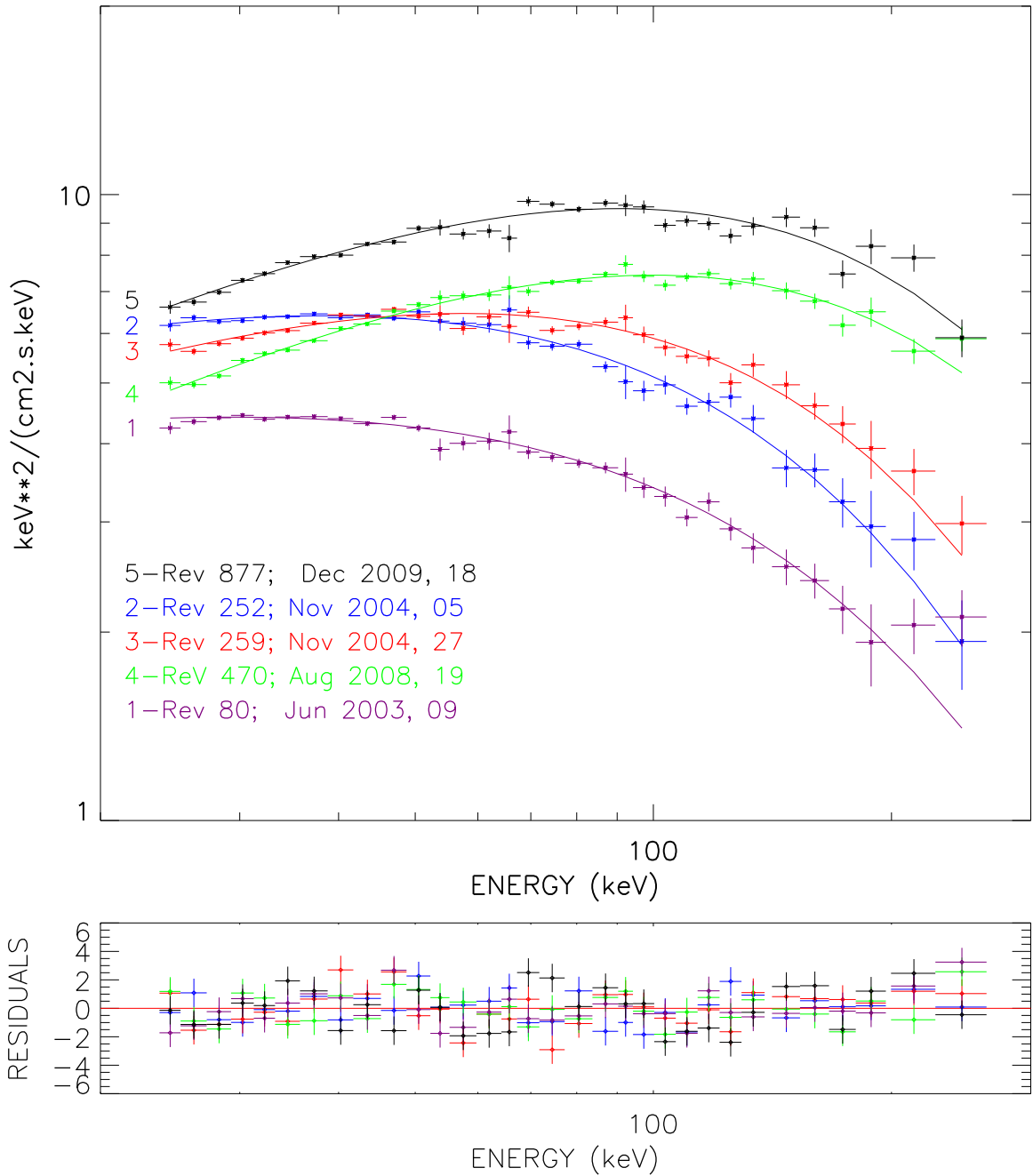} 
\caption{Individual  spectra illustrating the spectral evolution.
The 3 middle spectra illustrate the pivoting of the shape at constant flux, while
the lowest ($N^{0}$1) and the highest ($N^{0}$5) correspond to a change in intensity
with similar spectral shape.}
\label{fig:5spec.var}
\end{figure}

Conversely, it is worth noting that  different spectral shapes are observed for
 an unchanged intensity level in the 30-70 keV band. To illustrate both  of these effects, we have superimposed in Fig. \ref{fig:5spec.var}, five spectra representative of 
the global course followed by the source in Fig. \ref{fig:hrflux} (the corresponding revolutions 
are identified in this figure by squares, numbered from 1 to 5).

We recognize the two modes of evolution already identified
 for this source on the ks timescale (Malzac et al 2006):  a pivoting of the
spectral shape (spectra 2-4) and a change of luminosity  at constant hardness (spectra 1 \& 2
 and 4 \& 5). Note however, that the pivot
energy is here at $\sim$ 45 keV  (in the middle of the studied energy band, by construction).
It could be attributed to an increase of both temperature and optical depth of
the Comptonising medium.  When combined, the two variability modes give rise to a global complex behavior in the hard X-ray domain, although the source remains  in  the HS (and intermediate states).

To study in more details the spectral shape and its evolution in the high energy part, 
we have to accumulate data corresponding to the same hardness.

\subsection{Comparison of different averaged spectra }

\begin{figure}
\plotone{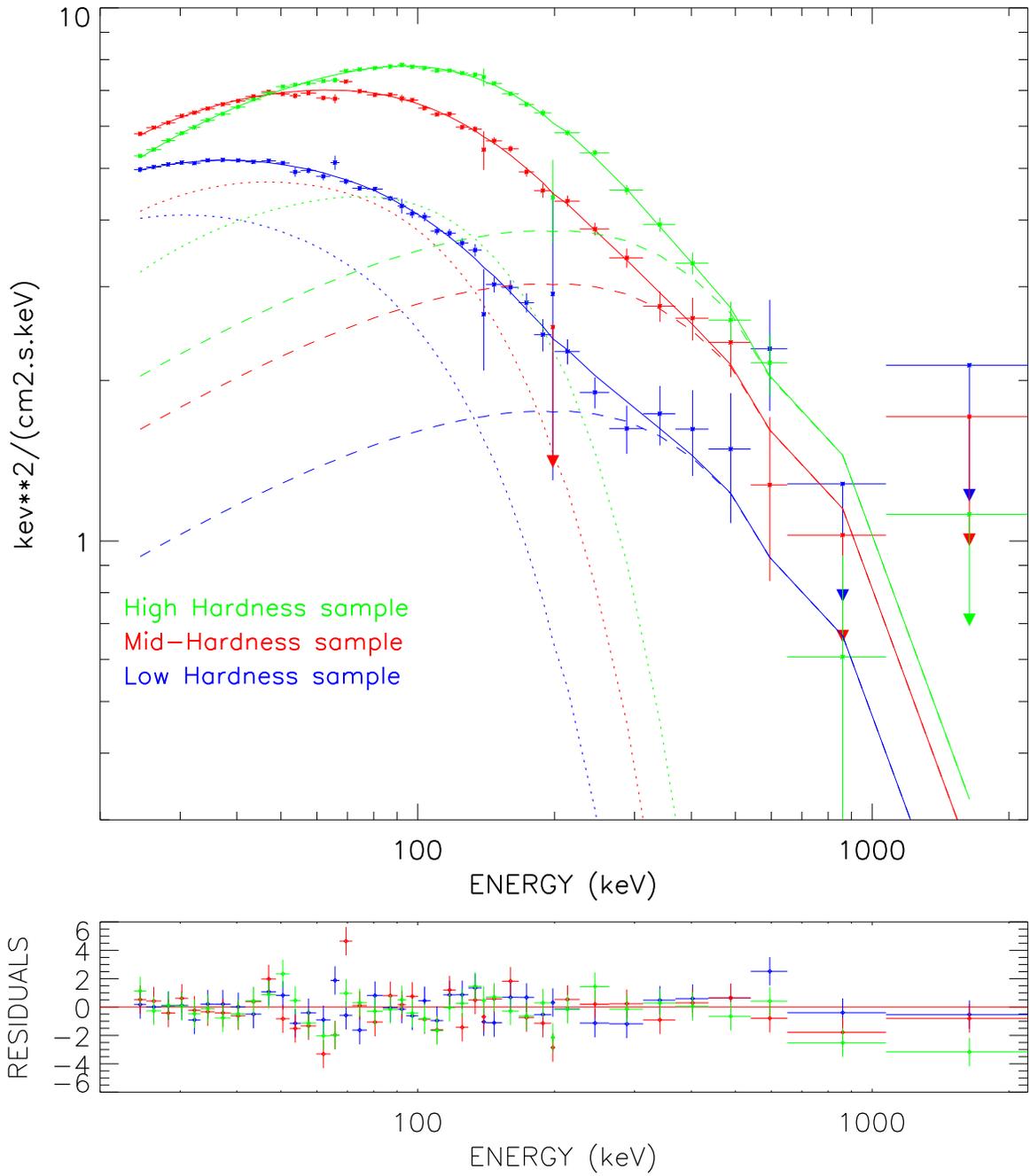} 
\caption{Cyg X-1 averaged spectra for the three hardness levels.
Solid lines represent  a model with 2 Comptonization models.
Dotted lines correspond to the first component (see Table 3 for the parameters) and 
dashed lines to  the second one (with kT fixed to 130 keV and $\tau$ fixed to 0.6).
Upper limits are at a 2 $\sigma$ level.  }
\label{fig:3HR}
\end{figure}

Based on the three hardness levels repered in Fig. \ref{fig:hrflux}, we have 
built the corresponding averaged spectra, hereafter refered to as 'low hardness', 'mid hardness'
and 'high hardness' samples, respectively. They are displayed on Fig. \ref{fig:3HR},
while best fit parameters are given in Table 3. Similarly to the total spectrum,
the data are modeled by a Comptonization law plus its reflection component 
(required when looking at residuals) and a second hotter Comptonization.
The evolution of the slope in the low energy part (in direct relation with the hardness)
is clear,  with an increase of the peak energy from $\sim$ 50 keV to $\sim$ 150 keV. 
This behavior is not related to the reflection component but appears as 
an evolution of the macroscopic parameters (kT and $\tau$) of the Comptonizing medium,
 which both increase from the soft to the hard sample (see Table 3).\\
Looking now to the high energy part, no significant emission is detected above 
700 keV. A degeneracy between parameters being unavoidable, 
we choose to fix the second Comptonizing medium temperature and optical depth to 130 keV and 0.6
respectively for all spectra, but free normalisation factors.\\
Even though it represents only a possible description of our data, 
this model, involving  a second (hotter and thiner) Comptonization
medium with constant parameters $kT_{e}$ and $\tau$, provides an acceptable interpretation
of the Cyg X-1 behavior in the hard X-ray domain.
Even if other models could reproduce the data, such two Comptonizing regions
(or more generally, variable $kT_{e}$ and/or $\tau$) scenario resembles those already applied
in previous works to Cyg X-1 and several BHBs (see for example the Suzaku observations
 analysed by Makishima et al, 2008, and references therein).
Beyond the specific sets of best fit parameter formally obtained, the inadequacy
 of the single comptonization region suggests  spatial and/or rapid temporal
  variation of the temperature or optical depth of the Comptonization region.

\section{Annihilation feature}
%
 Cygnus X-1 is one of the brightest Galactic sources of hard X-rays up to several
  hundreds of keV and is therefore a good candidate for positron production.  
The excellent energy resolution of the SPI germanium detector  makes it the best instrument to 
seek for potential annihilation signatures in the observed emission.\\
We have thus looked for any emission feature, narrow (10 keV FWHM) or broad (80 keV FWHM), 
transient (scw ie $\sim$ 2 ks timescale) or more persistent (revolution ie 1-2 day timescale).
On the scw timescale,  no significant emission is reported with 2 $\sigma$ upper limits of
2-3  $\times 10^{-3}$ \phs\ and 0.5-1 $\times 10^{-2}$ \phs\ for a narrow and broad line,
respectively. On the revolution timescale, no significant excess above
the continuum contribution are  found, with upper limits ranging between
  3-6 $\times 10^{-4}$ \phs\ and 0.7-1.1 $\times 10^{-3}$ \phs\, according to the revolution duration.
   Finally, considering the whole set of data (4 Ms),  persistent emission features,
if any, should be below 6 $\times 10^{-5}$ \phs\ and 1.3 $\times 10^{-4}$ \phs.


\section{Comparison with other instruments}

\begin{figure}
\plotone{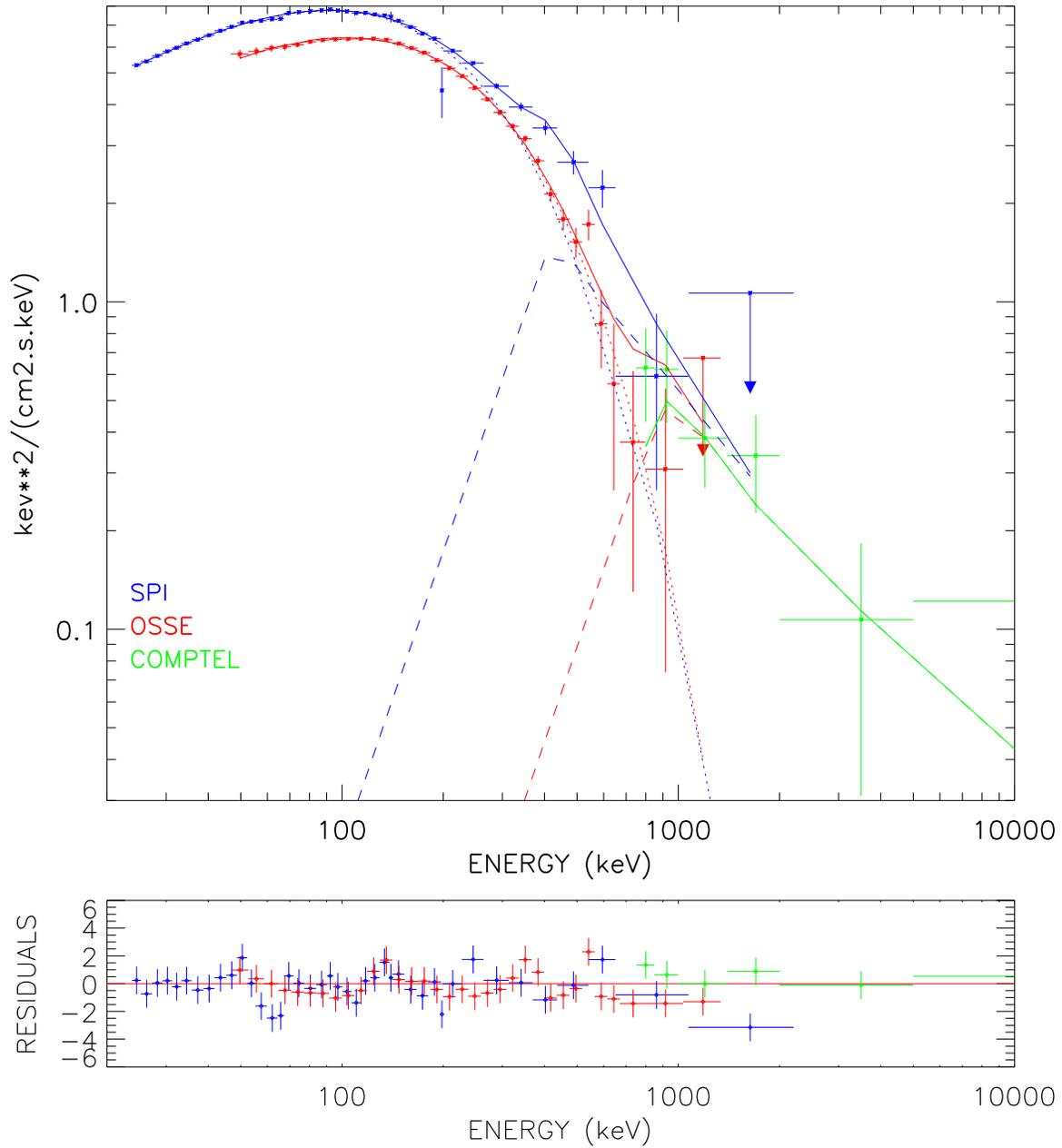}

\caption {Cyg X-1 composite spectrum  from CGRO (OSSE in red, COMPTEL in green, averaged Hard state,
from McConnell et al., 2002)
and INTEGRAL/SPI (in blue, high hardness sample averaged spectrum).  For each instrument,
the dotted curve corresponds to a cutoff power law component, the dashed curve 
to a broken power law  with first photon index fixed to -1, 
solid line to  the total (See Section 5).
}
 \label{fig:SPIGRO}
\end{figure}

We now compare our observations with the available data in the MeV region.
From Fig. \ref{fig:SPIGRO}, we can see that our "high hardness sample" observations  has a spectral shape that is very close to that of  the
 HS  mean spectrum reported by McConnell et al. (2002) from GRO/ COMPTEL and OSSE instruments.
 
To compare the whole set of data (INTEGRAL/SPI + CGRO/Comptel+OSSE), we use a model consisting in a cutoff power
law plus a broken power-law with the first index fixed to -1 ( to avoid any contribution
in the low energy part).\\
 Imposing the same photon index for the cutoff power-laws, the fit
procedure converges toward slightly different cutoff energies ($\sim$ 140 and 160 keV,  
 for SPI and OSSE data
respectively, no contribution in COMPTEL range).


We also impose a common second photon index in the broken power law  
and obtained 3.4 $\pm$ 0.5, while the energy break reflects the true difference between the spectra:
close to 900 ($\pm$ 100) keV in the CGRO data, it means that this component is based essentially on the
COMPTEL points. Around 420 ($\pm$ 25) keV in the SPI data, it allows us to 
recover the additional emission above the low energy component, the very 
soft index limiting its extension toward high energies. \\
In conclusion, even if the SPI data do not show significant emission above 1 MeV, the upper limits
are marginally consistent with the non-thermal tail observed at several MeV by CGRO/COMPTEL. We note however that this emission is not taken into account by the two-zone thermal 
Comptonization model proposed above which would require an additional component to fit the COMPTEL data. 
The overall good agreement between the SPI and OSSE spectra indicate that the average spectral
properties in the hard state have remained constant between the two epochs.

\begin{figure}
\plotone{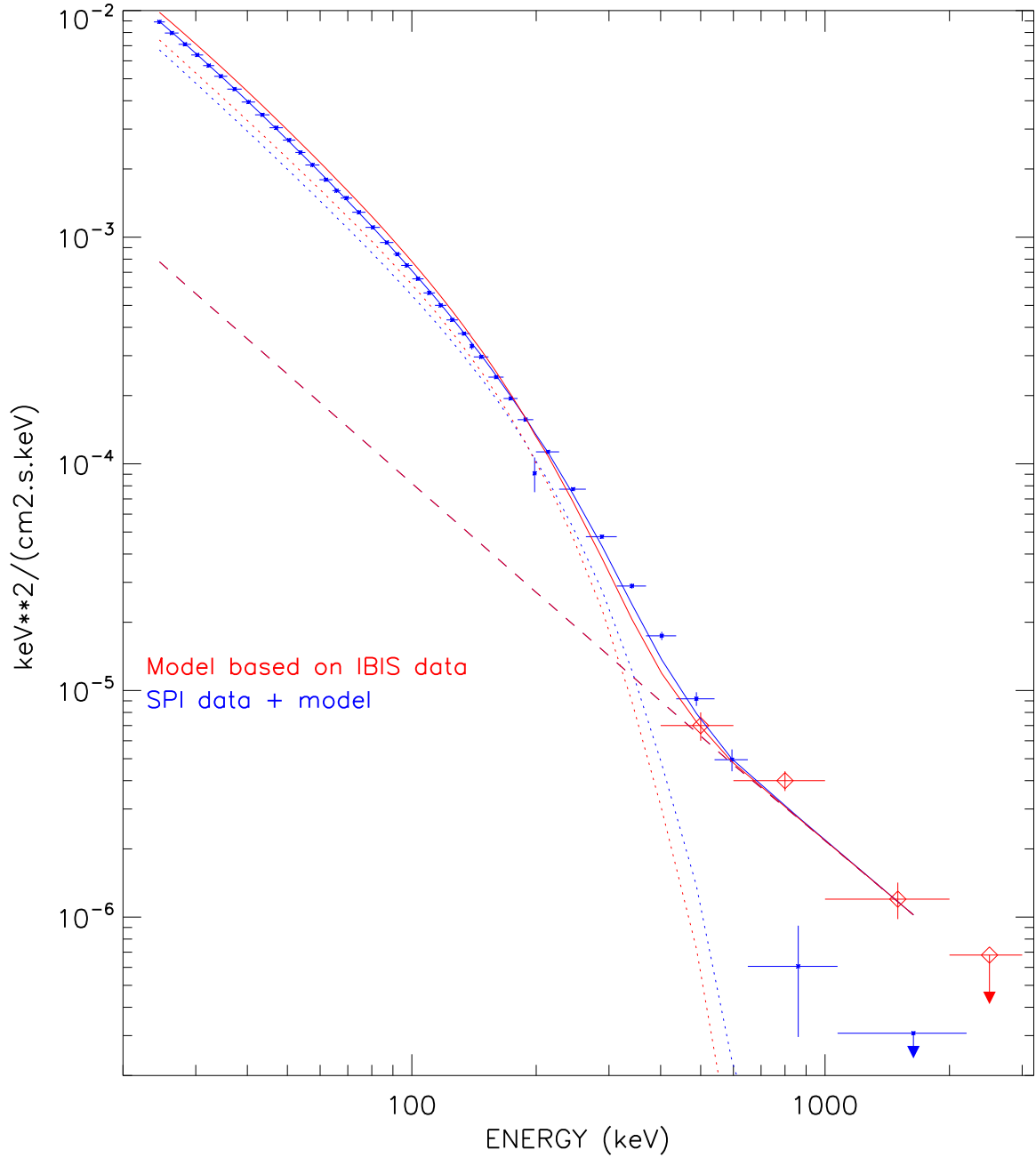}
\caption {Comparison the stacked spectra of SPI with that obtained by IBIS (Laurent et al. 2011). The IBIS data (red crosses) were extracted from Fig.~1 of Laurent et al. (2011) and then fit with a thermal 
Comptonization (CompTT, red dotted curve) plus power-law model (red dashed line). The SPI data and model are shown in blue. In the fit of the SPI data the powerlaw parameters were fixed at the same values as that of IBIS.  For clarity, the IBIS data are shown only above 400 keV.
}
 \label{fig:laurent}
\end{figure}

Recently, Laurent et al. (2011) presented a stacked spectrum of  Cygnus X-1 obtained
using the IBIS data during nearly the same observation period as ours.  IBIS is
composed of two position sensitive detector layers ISGRI (CdTe, 15-1000 keV; Lebrun
et al. 2003), and PICsIT (CsI, 200 keV-10 MeV; Labanti et al. 2003).  These two 
detectors are usually used independently to produce spectra and light curves. 
Laurent et al. (2011) used the IBIS/ISGRI data up to $\sim$ 400 keV  
and IBIS/PICsiT at 
higher energies. Fig.~\ref{fig:laurent} compares our averaged SPI  spectra with the 
results of Laurent et al. 2011.  
We note that most of the published results  from IBIS actually use only the ISGRI 
detector. With the recent versions of the public software, the results from IBIS/ISGRI 
are generally compatible with those of SPI. 
This  good agreement is confirmed by our comparison of the stacked SPI and IBIS/ISGRI 
spectra.
At higher energies however, the results  are clearly different: 
the SPI fluxes are  lower than the IBIS/PICsiT points by  a factor of about 5 at least. 
The origin of this disagreement is unclear. We were not able to go further in the 
analysis of the discrepancy as there are very few published results from IBIS/PICsiT.

\section{ Summary and Conclusions}
We used 4 Ms of \textit{INTEGRAL} SPI data  to study the emission of  Cyg X-1 from 20 keV 
to the MeV region. While the source has been essentially detected in the HS,
its presents  complex variability on all timescales.
 We have studied  the luminosity  and hardness 
evolution of the source  above 20 keV, on the scw (2000 s)  and revolution (1-2 days) 
timescales.   The revolution averaged data give a nice picture of the long term source behavior.
A change by a factor of 2-3 in luminosity can be accompanied either by a significant 
softening in the 20-150 keV  domain (see Fig. \ref{fig:3HR})
or by the same spectral shape just moved up and down.


Then, in order to improve the photon statistics at high energies we combined observations to produce long exposure time average spectra.  The analysis of the stacked spectra reveals that the  emission of Cyg X-1 extends up to $\sim$ 700 keV  
but presents a sharp cutoff around that energy. Whatever the criterion we used to
built averaged spectra  (low/mid/high hardness, low/high intensity, all data),
no emission can be detected above 1 MeV.\\
Nevertheless, a single Comptonization model does not provide a good description of
the overall spectral shape.  We have shown that a two temperature model provides a good fit
 of the SPI data, with a minimum of free parameters ($\tau$ and kT of the second Compton component can be considered as constant along the time).


In a final  step to investigate the high energy emission, we compare
all the data available above 300 keV for Cyg X-1 in the last 2 decades
and conclude that while our SPI upper limits are marginally 
compatible with the soft powerlaw reported by COMPTEL in the 90' (McConnell et al, 2002),
 they clearly disagree  (lower by a factor $\sim$ 5) 
with the recently reported IBIS/PICsiT emission (Laurent et al, 2011).\\
In other words, the presence  of a non-thermal mechanism
participating to the power supply can not be  excluded, but 
 our results contradict the presence of a hard (polarised or not) power law emission.

 A last point concerns the potential emission linked to the positron production:
No positive detection has been found, in a narrow (10 keV)  and broad (80 keV) channel,
on 2ks and day timescales, as well as in the total dataset.

The  question of the origin and the nature of the hard state 
high energy emission remains thus of prime interest. 
Cyg X-1 is clearly the most adequate target to investigate it.
Since no mission operating above 300 keV is expected for a while,
the only hope is that the INTEGRAL instruments will be able to accumulate a few more Ms
 of data on this important target to resolve the matter once and for all.

\section*{Acknowledgments}  The \textit{INTEGRAL} SPI project has been completed under the
  responsibility and leadership of CNES.
   We are grateful to ASI, CEA, CNES, DLR, ESA, INTA, NASA and OSTC for support.
   We thank A. A. Zdziarski for providing us with GRO/COMPTEL and OSSE data.


\clearpage
\begin{deluxetable}{lccccccc}
\tablewidth{0pt}
\tablecaption{Log of the \textit{INTEGRAL} SPI observations of Cyg X-1 used in this paper.}

\label{tab:revol}
\tabletypesize{\scriptsize}
\tablehead{
 \colhead{}\\
\colhead{revol }
&\colhead{Start} 
&\colhead{End} 
&\colhead{useful}\\
\colhead{number}
&&&\colhead{duration (ks)}\\
&\colhead{} \\
 }
\startdata

 79-80 (5x5)& 2003-06-07 00:59  &2003-06-12 03:35   & 293         \\
210-214 (A)& 2004-07-03 00:01 &2004-07-17 00:25  &  709        \\
251-252 (A)& 2004-11-03  14:23  & 2004-11-07 16:26   &  176   \\
259  \& 261 (H) & 2004-11-26 12:28 	 & 2004-12-03 15:43    &  143  \\
470 (EXO, H) &  2006-08-19 09:19 &2006-08-21 16:02  &  159    \\
486 (EXO, H) &  2006-10-06 00:11 & 2006-10-08 07:55  &  160  \\
498-505 (GP) &  2006-11-11 19:31 &2006-12-04 06:20   &   535      \\
628-631 (A) & 2007-12-04 19:05  &  2007-12-15 21:08  &  388     \\
673 (A) &  	2008-04-18 17:41  &2008-04-19 22:09   &  54      \\
682-684 (A)  & 2008-05-14 08:13 & 2008-05-22 19:54    & 304        \\
739-746 (A) & 2008-11-01 02:14 &2008-11-24 05:25 &  551    \\
803-806 (A) & 2009-05-11 08:27  &2009-05-22 11:32  &  371    \\
875(H*) \& 877(H)    & 2009-12-12 16:18  & 2009-12-19 20:57    & 160     \\

\enddata

\tablecomments{In the first column, the letter after the revolution number indicates the dithering strategy used:
(5x5) for the standart 5X5 pattern (see section 2); (A) for a pointing strategy centered between Cyg X-1 and Cyg A region;
(H) for the hexagonal pattern and (GP) for a Galactic Plane scan. (H*) During the rev 875, the pointing strategy 
follows a pattern proposed by Wilms et al in their AO-7 proposal. All this information
is available on the dedicated ESA site web http://integral.esa.int/isocweb.
}
 
\end{deluxetable}

\begin{deluxetable}{lcccccccc}
\tablewidth{0pt}
\tablecaption{Fit parameters for the total averaged spectra of Cyg X-1 presented in Fig. 2}
\label{tab:fittot}
\tabletypesize{\scriptsize}
\tablehead{
 \colhead{}\\
\colhead{Model }
&\colhead{$\Omega$} 
&\colhead{kT} 
&\colhead{$\tau $}
&\colhead{$\alpha$ or $ kT_{2}$}
 &\colhead{$\tau_{2} $}
&\colhead{$\chi_{red}^{2} $}\\
  &&\colhead{keV}
 &&\colhead{~~~~keV}
&&\colhead{(DoF)} \\
 }
\startdata  
Refl*Comptt&  0.90 $\pm$ 0.3  & 75.0 $\pm$ 3   & 0.91 $\pm$   0.03  &   & & 6.0 (37) \\
Refl*Comptt + power law&   0.88  $\pm$  0.3   &  56  $\pm$  3 &  1.2  $\pm$  0.1    & 1.8 $\pm$ 0.2  &  & 5.1 (35)  \\
Refl*Comptt+Comptt &  0.8 (fixed)	 &   38  $\pm$  3   &  1.6  $\pm$  0.15 & 123$\pm$   10&  0.5 (fixed)  &   1.7 (36)  \\
\enddata
\tablecomments{Parameters obtained for the mean Cyg X-1 spectrum  (all publicly available
 observations  corresponding to  4 Ms of useful duration). 0.5\% of systematic
 have been added to the data. Two parameters have been fixed in the second model
to overcome some degeneracy.
} 
\end{deluxetable}

\begin{deluxetable}{lcccccccccccccc}
\tablewidth{0pt}
\tablecaption{Fit parameters for 3 averaged spectra of Cyg X-1 (see Fig. \ref{fig:3HR})}
\label{tab:fit3hr}
\tabletypesize{\scriptsize}
\tablehead{
\colhead{}\\ 
\colhead{Sample/Model }
&\colhead{$\Omega$} 
&\colhead{kT} 
&\colhead{$\tau $}
&\colhead{$ kT_{2}$}
 &\colhead{$\tau_{2} $}
&\colhead{$\chi_{red}^{2} $}\\
  &&\colhead{keV}
 &&\colhead{keV}
%
 }
\startdata 
 Low hardness sample & 0.55 $\pm$ 0.25 &  33 $\pm$ 3  & 1.2 $\pm$ 0.1   & 130 (fix)    & 0.6 (fix)  &  0.99 (36) &   \\
 Mid hardness sample & 0.73 $\pm$ 0.15 &36  $\pm$ 3  & 1.4 $\pm$ 0.2    & 130 (fix)    & 0.6 (fix) &2.13 (36)    \\
 High hardness  sample  & 0.68 $\pm$ 0.15 &  40 $\pm$ 3  & 1.6 $\pm$ 0.1   & 130 (fix)    & 0.6 (fix)  &  1.4 (36) &   \\
\enddata
\tablecomments{Best fit parameters for the three mean Cyg X-1 spectra (grouped by same hardness).
The model consists in two Comptonization components, the second one
with fixed parameters (except normalisation). 0.5\% of systematic
have been added to the data. 
}
\end{deluxetable}     


\clearpage

\end{document}